\documentclass[
reprint,
superscriptaddress,
showpacs,
showkeys,
 amsmath,amssymb,
 aps,
 prl,
]{revtex4-2}

\usepackage{xcolor}
\usepackage{version}
\usepackage[T1]{fontenc}% Custom fix to enable small caps in bold font
\usepackage{graphicx}% Include figure files
\usepackage{dcolumn}% Align table columns on decimal point
\usepackage{bm}% bold math
\usepackage{hyperref}% add hypertext capabilities
\hypersetup{
    colorlinks=true,
    linkcolor=magenta,
    filecolor=red,      
    urlcolor=blue,
    pdftitle={},
}
\usepackage[all]{hypcap}
\usepackage[mathlines,modulo]{lineno}% Enable numbering of text and display math
\usepackage[inter-unit-product =\cdot]{siunitx}

\begin{document}

%\title{Anomalies in reactor antineutrino spectra in light of a new summation model with parameterized missing transitions}

\title{On the origin of the reactor antineutrino anomalies in light of a new summation model with parameterized $\beta^{-}$ transitions}

\newcommand{\IRFU}{\affiliation{IRFU, CEA, Universit\'e Paris-Saclay, 91191 Gif-sur-Yvette, France}}
\newcommand{\LNHB}{\affiliation{Universit\'e Paris-Saclay, CEA, List, Laboratoire National Henri Becquerel (LNE-LNHB), F-91120 Palaiseau, France}}

\author{A.~Letourneau}\email{alain.letourneau@cea.fr}\IRFU
\author{V.~Savu}\IRFU
\author{D.~Lhuillier}\IRFU
\author{T.~Lasserre}\IRFU
\author{T.~Materna}\IRFU
\author{G.~Mention}\IRFU
\author{X.~Mougeot}\LNHB
\author{A.~Onillon}\altaffiliation{Present address: Technical University of Munich, JamesFranckStrasse 1, Garching, 85748, Germany}\IRFU
\author{L.~Perisse}\IRFU
\author{M.~Vivier}\IRFU

%\date{\today}

\begin{abstract}

We investigate the possible origins of the norm and shape reactor antineutrino anomalies in the framework of a summation model (SM) where $\beta^{-}$ transitions are simulated by a phenomenological Gamow-Teller $\beta$-decay strength model. The general trends of the discrepancies to the Huber-Mueller model on the antineutrino side can be reproduced both in norm and shape. From the exact electron-antineutrino correspondence of the SM model, we predict similar distortions in the electron spectra, suggesting that biases on the reference fission-electron spectra could be at the origin of the anomalies.

\end{abstract}

\pacs{14.60.Lm, 14.60.Pq, 14.60.St, 28.41.--i}% PACS, the Physics and Astronomy Classification Scheme.
\keywords{Neutrino Flux, Nuclear Reactor, Sterile Neutrinos, Uranium}%Use showkeys class option if keyword display desired

\maketitle

The reactor antineutrino anomalies are a several-years long standing problem in neutrino physics. They refer to an observed $\sim$6\% deficit in the detected rate, known as "Reactor Antineutrino Anomaly" (RAA), and a $\sim$10\% excess of events in the 4-6 MeV range, known as the "5-MeV bump", when comparing experimental data to the prediction of the state-of-the-art Huber-Mueller (HM) model \cite{HUBER,MULLER}. The RAA was first put in evidence in \cite{MENTION} by comparing to short baseline reactor experiments, and confirmed by all recent high precision reactor antineutrino experiments at distances of 300-500 m from the reactor \cite{DAYABAY-RAA,DOUBLECHOOZ-RAA,RENO-RAA} and below 30 m \cite{STEREO-RAA}. The "5-MeV bump" is observed in all the above-cited high precision reactor antineutrino experiments although with slightly different amplitudes and shapes \cite{DAYABAY-Bump,RENO-Bump,DOUBLECHOOZ-Nature,NEOS-Bump,STEREO-Bump,PROSPECT-Bump}.

At present, no consensus has been reached concerning the origins of these anomalies. The RAA was first interpreted as the possibility of the existence of a hypothetical sterile neutrino state, mixing with the active electronic flavor.
The best fit parameters for this sterile state to absorb the anomaly was found around  1 eV$^{2}$ for the oscillation frequency ($\Delta m^{2}$) and 0.14 for the amplitude ($sin^{2}2\theta$) \cite{MENTION}. This best fit region of oscillation parameters is now rejected to high confidence level by several experiments \cite{NEOS-Bump, STEREO-Sterile,PROSPECT-Sterile,DANSS-Sterile, KATRIN-Sterile} that have tested the sterile neutrino hypothesis in a model independent way.

On the other hand, the Daya Bay and RENO experiments \cite{DAYABAY-Isotopic, RENO-Isotopic} have studied the dependence of the antineutrino yield to the fuel-composition. They concluded that a $\sim$8\% bias of the $^{235}$U Inverse Beta Decay (IBD) yield could be solely responsible for the RAA. This result is slightly in tension with experiments at research reactors with pure $^{235}$U fuel showing a (5.0$\pm$1.3) \% deficit \cite{STEREO-RAA}, not allowing to conclude.
But the hypothesis of a normalization bias on $^{235}$U spectrum is reinforced by the recent measurement of the $^{235}$U to $^{239}$Pu electron energy spectra ratio \cite{KOPEIKIN} reporting a constant $\sim$5\% disagreement with respect to the HM prediction.

Regarding the shape anomaly, extensive studies \cite{DWYER,FANG,HAYES-2015,SONZOGNI-2016,SONZOGNI-2017} have been conducted to find explanations in the prediction modeling but none of them have succeeded to bring satisfactory solutions.

The Huber-Mueller model is based on an improved method to convert the cumulative $\beta^{-}$ spectra measured at ILL with the BILL spectrometer \cite{BILL-U235,BILL-Pu239,BILL-Pu241} into antineutrino spectra. In this method, if experimental biases exist on the measured $\beta$ spectra, they would be transferred to the converted antineutrino spectra and could be at the origin of the anomalies.

The present contribution proposes to use the exact electron-antineutrino correspondence of a refined summation model (SM) to test the consistency of the electron and antineutrino spectra predicted by the HM model and to search for biases in the original $\beta^{-}$ spectra used as reference to construct the HM model.

The summation method consists of treating the electron and antineutrino energy spectra at the level of single $\beta$-transitions and to sum over all the transitions for all the decaying fission fragments. Thus electron and antineutrino spectra are calculated within the same theoretical framework, using the same inputs, preserving the symmetry of the two leptons spectra at the single-branch level. The cumulative energy spectrum for electrons or antineutrinos produced in a reactor writes:

\begin{equation}
    S_{f}(E,t) = \sum_{f} A_{f}(t) \sum_{b} I_{b} \times S^{b}_{f}(E)
    \label{eq:neutrino flux}
\end{equation}  where the $f$ index runs over all the fission fragments, the $b$ index runs over all the $\beta$-branches for one fission fragment. The time-dependent term $A_{f}(t)$ is the $\beta^{-}$-activity of the fragment after a time $t$ of irradiation. We used the FISPACT-II code \cite{FISPACT} with the JEFF3.3 \cite{JEFF} independent fission yields as input to compute the activities after the short irradiation times used for the reference ILL measurements. This approach allows to account for non-equilibrium effects compared to the use of cumulative yields. 

The energy dependent term $S_{f}(E)$ is the energy spectrum of the electron or antineutrino corresponding to one transition. We used the BESTIOLE code developed in \cite{MULLER} and recently updated in \cite{PERISSE} to calculate this term. It integrates the most sophisticated first-principle derivations of the Fermi theory with finite size and weak magnetism corrections. For simplification and to highlight the effect of missing transitions, only allowed transitions are considered in this work.

Finally, the $I_{b}$ term is the intensity of the transition to the excited state $E_{j}$ of the daughter nucleus. In this work, $I_{b}$ is derived from a Gamow-Teller strength $B_{GT}(E_{j})$ model that will be detailed later, using:

\begin{equation}
    I(E_{j})= \frac{f(E_{j})T_{1/2}}{K}B_{GT}(E_{j})\lambda^{2}
    \label{eq:GT beta-strength}
\end{equation}  where $K = 6143.6(17)$ \cite{HARDY_Kvalue}, $\lambda=g_{A}/g_{V}=-1.270(3)$ \cite{HARDY_lambda}, $T_{1/2}$ is the half-live of the isotope. The phase-space factor is calculated as:

\begin{equation}
	f(E_{j}) = \int_{0}^{Q_{\beta}-E_{j}}  \mathcal{F}(Z,A,E_{j}) \sqrt{E(E+2m_{e}c^{2})} E(E_{j}-E)dE
\end{equation}  where $\mathcal{F}(Z,A,E_{j})$ is the usual Fermi function for the daughter nucleus.
Fermi transitions are neglected in this work because their strength is contained within a very narrow resonance located above the $\beta^{-}$-decay window for most of the fission fragments.
 
The SM fully relies on the available data or models for the fission yields and for the $\beta$ decay. As stated in previous studies \cite{MULLER,HAYES-2015, SONZOGNI-2016,FALLOT,ESTIENNE} it suffers from the uncertainties or lack of data in the databases. Whereas fission yields for the two major actinides $^{235}$U and $^{239}$Pu are quite well known, the situation is different for $\beta$-decay data. The ENSDF $\beta^{-}$-decay database \cite{ENSDF} suffers from a lack of knowledge for the more unstable isotopes and, when they are known, some transition intensities are affected by the Pandemonium effect \cite{HARDY}. It results in an underestimation of the total number of electrons and antineutrinos per fission and an overestimation of the high-energy part of the reconstructed energy spectra \cite{MULLER}. To circumvent this problem, the most sophisticated summation model \cite{FALLOT, ESTIENNE} integrates experimental data from calorimetric measurements and the Gross theory to complete the database at high excitation energy. Unfortunately, calorimetric measurements do not cover all the fission fragments that need to be corrected and the problem of missing transitions still persists. 

We then have developed a $\beta$-decay strength model to generate missing transitions for all fission fragments and to complete the ENSDF database. The Gamow-Teller strength model proposed in this work is a phenomenological model based on an exhaustive analysis of the experimental $\beta^{-}$-decay strengths extracted from the low-resolution Total Absorption Gamma Spectrometry (TAGS) measurements \cite{GREENWOOD,FALLOT,RICE,GUADILLA-2019,GUADILLA-2017,JORDAN} using the inverse of Eq.\ref{eq:GT beta-strength} (see \cite{SAVU} for more details on the model). The strengths exhibit an universal feature as a function of the excitation energy: a discrete domain below 2-3 MeV and a continuous domain above, with a resonance structure. The properties (resonance spacing, width and amplitude) were extracted by fitting the TAGS strengths with Gaussian distributions. The resonance spacing varies between 0.1 and 1 MeV with a mean spacing centered around 340 keV. The resonance widths slightly depend on the excitation energy and vary between 20 and 100 keV. The resonance amplitudes are contained within an envelope for which the upper limit follows the nuclear level density trend.
All these properties are used to generate randomly resonances considering a Gaussian shape for each resonance. 
The nuclear level densities calculated by Hartree Fock Bogolyubov (HFB) technique plus combinatorial deformations based on the BSk14 Skyrme force from RIPL3 \cite{RIPL3} were used. A global scaling factor $N_{0}$ was applied to the densities to match with the experimental strength values and a quenching factor $\alpha$ was added to reduce an overestimation of the calculated nuclear level density as suggested in \cite{HFB-corr}. The upper limit for the amplitudes as a function of the excitation energy ($E^{*}$) then writes:

\begin{equation}
    B_{GT}^{up}(E^{*}) = N_{0}e^{-\alpha \sqrt{E^{*}}} \rho_{HFB}(E^{*})
\end{equation}

\begin{figure}[h]
    \includegraphics[width=1.\columnwidth]{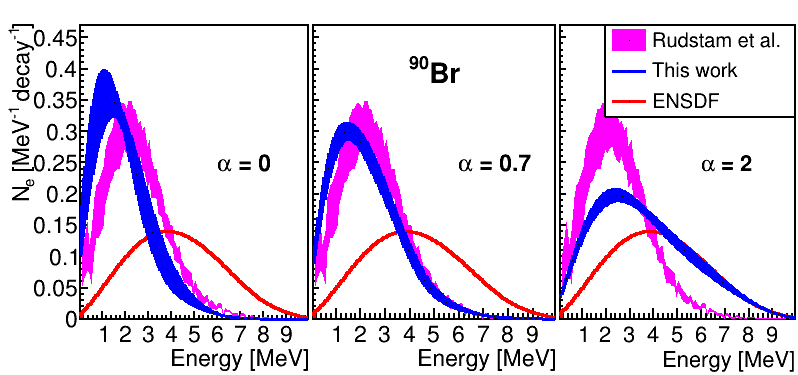}
    \includegraphics[width=1.\columnwidth]{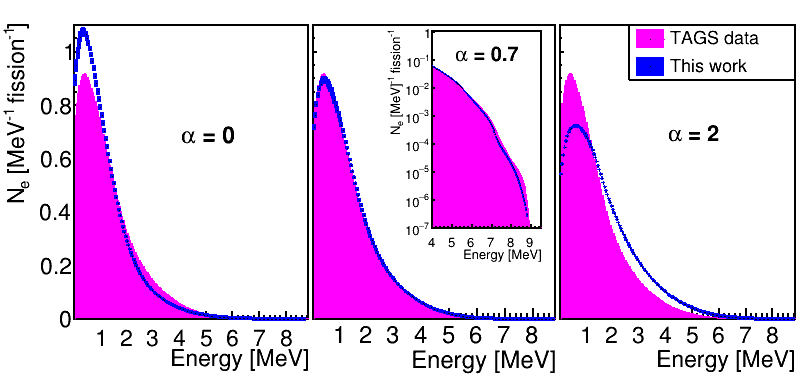}
    \caption{Top: electron kinetic energy spectra for $^{90}$Br generated with the summation model using the $\beta$-strength model (in blue) with three $\alpha$ values and the ENSDF-2020 nuclear databases (in red) as inputs and compared to the experimental spectrum from \cite{RUDSTAM}. Bottom: cumulative electron kinetic energy spectra calculated with the summation model using the GT strength model as input (in blue) or the experimental TAGS data (in magenta) for the three same $\alpha$ values. The summation was done using $^{235}$U fission yields, considering only isotopes measured in TAGS experiments.}
    \label{fig:Tengblad_electron}
\end{figure}

The $\alpha$ parameter taken as common to all nuclei in this simple approach, is the only free parameter of the model. Such an approach allows an overall good reproduction of the $\beta^{-}$ energy spectra (see top part of Fig.\ref{fig:Tengblad_electron} for example) but does not allow to reproduce perfectly all the individual energy spectra with the same $\alpha$ parameter.

To preserve the structure of the $\beta$-decay strength at low excitation energy, the discrete region is filled using the data from the $\beta^{-}$-decay ENSDF database \cite{ENSDF} when they exist (for about 96.5\% of the total fission fragments activity for $^{235}$U after 12h of irradiation). In that case, the ENSDF intensities are used in a relative way. This allows to correct them after adding the continuous part of the $\beta$-decay strength. When no data exists, discrete transitions are generated randomly following the HFB level density, the Q$_{beta}$ value being extracted from the calculated table of masses NUBASE \cite{NUBASE}. The energy cut-off between the discrete and the continuous domain is determined as the energy above which the HFB level density is higher than 100 MeV$^{-1}$.

\begin{figure*}[ht]
    \centering
    \includegraphics[scale=0.42]{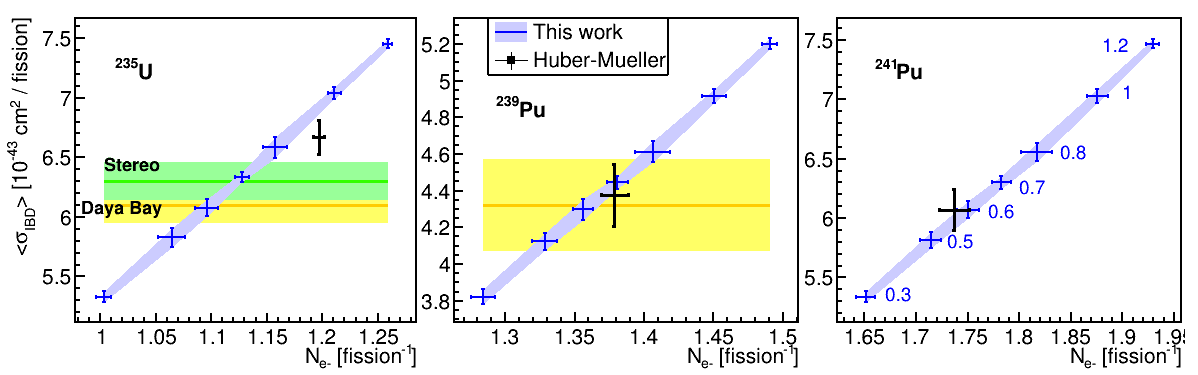}
    \caption{Average IBD cross section per fission versus the number of electrons per fission calculated with the $\beta$-strength model for the different $\alpha$ values labeled in the right plot for the three fissioning systems. The band results from the uncorrelated uncertainties due to the $\beta$-strength model. Correlated uncertainties on the fission yields are not added. They amount to 0.7\%, 0.9\% and 2\% for $^{235}$U, $^{239}$Pu and $^{241}$Pu, respectively. The cross lines represent the Huber-Mueller model. Experimental results (color bands) from Daya Bay \cite{DAYABAY-Isotopic} and STEREO \cite{STEREO-RAA} are also indicated. The same range of integration than Huber-Mueller was used for $N_{e}$.}
    \label{fig:Rate comparison}
\end{figure*}

Uncertainties introduced by the stochastic nature of the strength model are calculated by running about 100 times the summation model. At each iteration a new strength is generated for each isotope.

The strength model was first validated by comparing the electron energy spectra calculated with our summation method with the existing measured energy spectra from Rudstam et al. \cite{RUDSTAM}. An example is shown on Fig.\ref{fig:Tengblad_electron} (top) for $^{90}$Br, a large Q$_{\beta}$ isotope with major contribution of missing transitions. Three values of $\alpha$ are plotted defining a range of physical meaning for the model. The improvement of the GT strength model with respect to pure ENSDF inputs is clearly visible. The free parameter $\alpha$ allows to tune the induced correction for an optimal agreement between our prediction and the experimental spectrum, here obtained for $\alpha \sim 0.7$.

Then the model was validated by comparing to the cumulative electron energy spectra calculated using TAGS data as input (see Fig.\ref{fig:Tengblad_electron} bottom). The best agreement is also observed for $\alpha \sim 0.7$. Note that the high energy part of the spectrum is not well reproduced due to the dominant contribution of $^{92}$Rb in this region. This isotope is a particular case with a quasi single transition to the ground state that our model is not able to perfectly catch: the transition intensity to the ground state is reduced by about 20\% in our model compared to the measured TAGS data. 
Note also, that the same agreement when comparing with cumulative antineutrino energy spectra using TAGS data as input is observed for $\alpha=0.7$.

\begin{figure*}[ht]
    \includegraphics[width=2.\columnwidth]{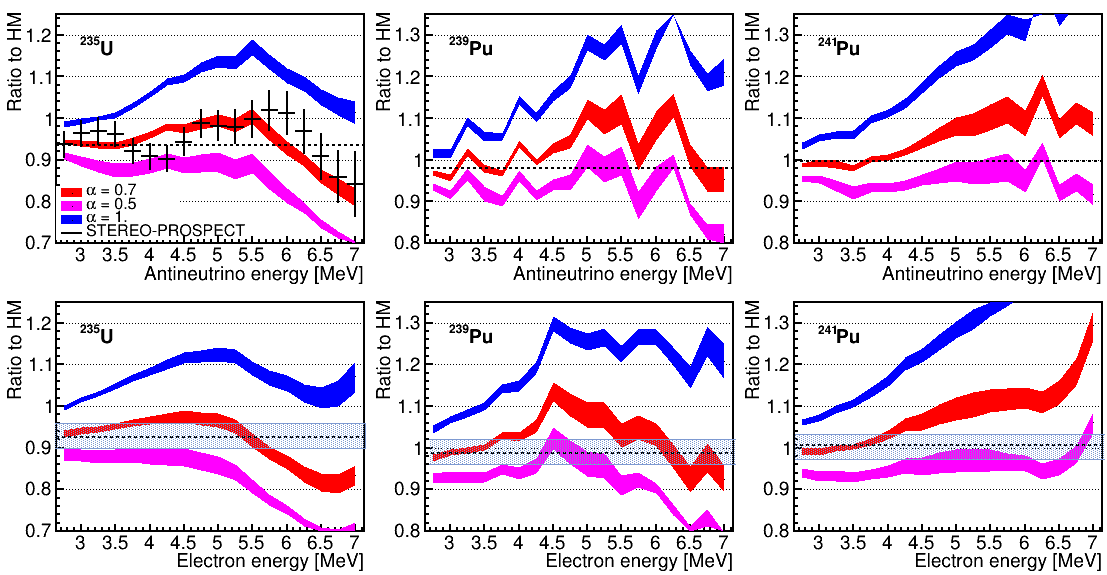}
	\caption{Ratios of the kinetic energy spectra for $^{235}$U, $^{239}$Pu and $^{241}$Pu, calculated with our model for different $\alpha$ values, to the Huber-Mueller model for antineutrinos (top) and electrons (bottom). The width of the lines indicates the standard deviation of the GT strength model due to the stochastic process. Uncertainties on the fission yields are not added but they have similar trend and amplitudes (see \cite{SupMaterial}). For comparison, the ratio to HM (scaled by a factor 0.95 \cite{STEREO-RAA}) constructed with the unfolded antineutrino spectrum from \cite{STEREO-PROSPECT} is added. The grey band on the electron side corresponds to the 90\% C.L. systematic uncertainties on the determination of the BILL spectrometer efficiency.}
	\label{fig:Comparison_with_ratio}
\end{figure*}

The comparisons with the HM model are discussed in the following for $^{235}$U, $^{239}$Pu and $^{241}$Pu fissioning systems (the model data with uncertainties for three values of $\alpha$ around 0.7 are available in \cite{SupMaterial}).

Figure \ref{fig:Rate comparison} shows the correlations, when varying $\alpha$, between the number of electrons integrated over the HM range, normalised per fission, and the average IBD cross section defined as:

\begin{equation}
    <\sigma_{IBD}> = \int\displaylimits_{\SI{2.}{MeV}}^{\SI{8.0}{MeV}} S_{f}(E_\nu) \sigma_\text{IBD}(E_\nu) \mathrm{d}E_\nu
    \label{eq.crosssection}
\end{equation}  using the IBD cross section from \cite{MENTION}. The use of this quantity rather than the number of neutrinos per fission is intended for direct comparison with the measured rates.

The HM model is found compatible with the alignment of our model for the three isotopes, suggesting that the bias of the conversion method, if it exists, is small and not sufficient to explain the RAA. The HM model is in agreement with our model for two different ranges of $\alpha$ values: between 0.8 and 0.9 for $^{235}$U and between 0.6 and 0.7 for $^{239}$Pu and $^{241}$Pu. As seen on Fig.\ref{fig:Rate comparison}, the latter range is also favored by STEREO and Daya Bay for $^{235}$U and $^{239}$Pu. 
This leads us to the conclusion that the RAA could find its origin in the overestimation of the $\beta^{-}$ spectrum measured after 12h of irradiation \cite{BILL-U235}.

Figure \ref{fig:Comparison_with_ratio} shows the ratios of the electron and antineutrino energy spectra calculated with our model to the HM model, for three $\alpha$ values around 0.7. On the antineutrino side, shape discrepancies clearly appear for all fissioning isotopes. The discrepancies have the form of a bump for $^{235}$U and $^{239}$Pu and a global linear deviation for $^{241}$Pu which does not depend too much on the $\alpha$ values. In the 4-6 MeV the excess of event around 5.5 MeV amount to 6\% for $^{235}$U, 12\% for $^{239}$Pu and 11\% for $^{241}$Pu when choosing $\alpha$=0.7. These values for $^{235}$U and $^{239}$Pu are in the vicinity of the ~10\% deviations observed in the recent experiments \cite{DAYABAY-Bump,RENO-Bump,DOUBLECHOOZ-RAA,NEOS-Bump,STEREO-Bump,STEREO-PROSPECT, PROSPECT-Bump, DAYABAY-Nuspectra}. As observed on the figure, the agreement with the ratio to HM model constructed with the unfolded antineutrino spectrum measured in the two pure $^{235}$U STEREO and PROSPECT experiments \cite{STEREO-PROSPECT} is very good ($\chi^{2}/ndf$=17/18) for $\alpha$=0.7. 
 Thus, with a simple model able to roughly reproduce and extrapolate TAGS data, with a single free parameter, our summation model is able to describe the deviations with respect to HM prediction measured by recent antineutrino experiments. 

All the distortions predicted on the antineutrino side are also expected on the electron side, with more or less the same amplitudes. Parts of them are contained within the systematic uncertainties of the BILL spectrometer efficiency determined from the dispersion of the (n,e$^{-}$) calibration reactions, but not all.

\begin{figure}
    \centering
    \includegraphics[scale=0.27]{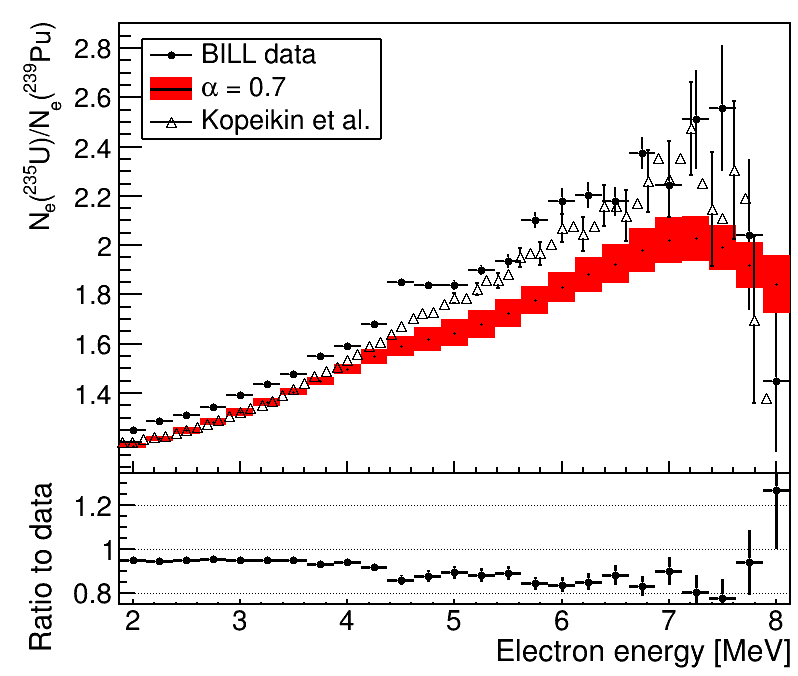}
    \caption{Ratios of the electron kinetic energy spectra for $^{235}$U and $^{239}$Pu calculated with our model for $\alpha$=0.7 and compared with experimental data from ILL \cite{BILL-U235,BILL-Pu239}. The uncertainties on the fission yields are not included. For comparison, the data from Kopeikin et al. \cite{KOPEIKIN} are added.}
    \label{fig:beta_spectra_ratios}
\end{figure}

As shown on Fig. \ref{fig:beta_spectra_ratios}, no further distortions are introduced on the $^{235}$U/$^{239}$Pu ratio and we confirm the $\sim$5\% of discrepancy observed by Kopeikin et al. \cite{KOPEIKIN}. The model is in quite good agreement with Kopeikin's measurement except above 4 MeV where small deviations appear, but uncertainties are larger in that region both in the experimental data and in the model.

All these different points lead us to the conclusion that the anomalous shape could be attributed to a shape bias in the $\beta^{-}$ energy spectra measured at ILL.

In summary, we have presented a phenomenological Gamow-Teller strength model able to simulate $\beta^{-}$-decay transition-intensities for fission fragments and to correct for Pandemonium effect and missing transitions in the ENSDF database.
Despite the simplicity of the model, the main features and divergences observed in antineutrino experiments compared to the Huber-Mueller model can be reproduced by a summation model with tuned input parameters. It highlights the importance of missing transitions in the modeling of antineutrino fission spectra.
Using the exact correspondence between electron and antineutrino in the summation approach, we have seen that equivalent deviations are expected on the electron side. The conclusions of this study suggest that the reactor antineutrino anomalies could find their origin in a norm bias for the measured $^{235}$U spectrum after 12h of irradiation and a shape bias for all measured electron spectra. Although these conclusions are supported by independent measurements, the origin of the biases are still unclear at this stage. Some biases on the neutron cross sections used to normalize the $\beta^{-}$ spectra could cover part of the RAA \cite{ONILLON} and part of the shape anomaly could be  included in the envelope of systematic of the BILL spectrometer efficiency. This work tends to confirm the need for improving the accuracy of $\beta^{-}$ fission spectra both on the experimental and theoretical sides.

\end{document}